\documentclass[twocolumn,showpacs,prl,amsmath,amssymb,floatfix]{revtex4}
\input{epsf}

\usepackage{graphicx}
\usepackage{dcolumn}
\usepackage{bm}

\begin{document}

\preprint{APS/123-QED}

\title{Vacuum Fluctuations induced Entanglement between Two Mesoscopic Systems}
\author{H. T. Ng and K. Burnett}
\affiliation{{Clarendon Laboratory, Department of Physics,
University of Oxford, Parks Road, Oxford OX1 3PU, United Kingdom}}
\date{\today}

\begin{abstract}
We study the dynamics of a pair of molecular ensembles trapped
inside a superconducting resonator through which they are strongly
coupled via a microwave field mode. We find that entanglement can be
generated via ``vacuum fluctuations'' even when the molecules and
cavity field are initially prepared in their ground state.  This
entanglement is created in a relatively short time and without the
need for further manipulation of the system. It does, therefore,
provide a convenient scheme to entangle two mesoscopic systems, and
may well be useful quantum information processing.
\end{abstract}


\maketitle

Vacuum fluctuations can have important physical consequences, for
example, in the Casimir effect \cite{Casimir} and Hawking radiation
\cite{Hawking}.  van der Waals interactions, i.e., attractive
long-ranged interaction between neutral atoms or molecules, are also
a kind of Casimir effect.  It is an interesting question as to how
vacuum fluctuations can be used to influence the properties of
quantum entanglement between two systems. Quantum entanglement is a
fundamental concept in quantum mechanics \cite{Einstein} and is also
the physical resource in quantum information processing
\cite{Nielsen}. In fact, it has recently been shown possible to
generate entanglement between a pair of particles via the vacuum
modes of the radiation field \cite{Reznik,Retzker,Cirone}.

In this paper, we study how vacuum fluctuations induce quantum
entanglement between two mesoscopic systems, i.e., polar molecular
ensembles \cite{Andre} are placed inside a cavity, and strongly
coupled by a single microwave mode. Recently, Rabl {\it et al.}
\cite{Rabl} have proposed the realization of a quantum memory using
such ensembles of polar molecules inside a superconducting resonator
\cite{Wallraff}.  The energy difference between two internal states
of a polar molecule is the order of GHz and polar molecules have
significant electric dipole moments. A strong coupling to a
microwave field via a transmission line can thus be achieved.  In
addition to the strength of the coupling, low-lying collective
excitations can be coupled to the field and exploit the enhanced
coupling to them, which scales as $\sqrt{N}$, where $N$ is the
number of molecules in the ensemble.

The dynamics of vacuum fluctuations \cite{remark1} is hard to
observe in ordinary systems.  To show why this is so, we start the
simple case of a two-level atom interacting with a quantized field.
Conventionally, we use the Jaynes-Cummings model \cite{Scully} in
the interaction picture, Hamiltonian
$H=g'[b^\dag\sigma_-e^{i(\omega'-\omega_0')t}+\sigma_+be^{-i(\omega'-\omega_0')t}]$,
to describe a two-level system $\sigma_\pm$ coupled to a quantized
field $b$, for $\omega_0'$, $\omega'$ and $g'$ are an energy
difference between two-level atom,  the frequency of the field and
the Rabi frequency respectively.  The rotating wave approximation
(RWA) can usually be used because the two countering-rotating terms,
$b\sigma_-$ and $\sigma_+b^\dag$, can be neglected; they carry a
fast oscillation with the high frequency $\omega'+\omega_0'$. The
RWA is, therefore, an excellent approximation for the optical
frequency regime in the weak Rabi coupling limit.  Clearly, this
Hamiltonian will produce no evolution in the atoms and the photon
field if they both start in the irrespective ground states. However,
this approximation breaks down if the Rabi frequency $g'$ is
comparable to the frequencies $\omega'$ and $\omega_0'$.   In fact,
the RWA is completely inadequate to describe the physical situation
of a large number of molecules interacting with a microwave field in
the strong coupling regime. It is thus necessary to go beyond RWA
and, in essence, study the role of vacuum modes on the dynamics of
the coupled atom-field system.

\begin{figure}[ht]
\includegraphics[height=2.5cm]{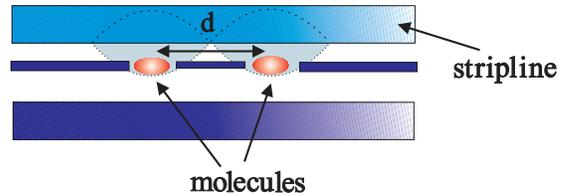}
\caption{\label{fig1} Two molecular ensembles, separate with a
distance $d$, are trapped inside a superconducting resonator and
interact with a single-mode microwave field along the stripline.}
\end{figure}

We consider the case where molecules and the photon field are
initially prepared in their ground states and show how the
countering-rotating terms in the Hamiltonian do indeed induce
quantum correlations between the molecules.   The two ensembles of
molecules exchange energy with the vacuum field due to the counter
rotating terms. In physical terms we would say that this proceeds
via virtual excitations of the cavity mode. We should bear in mind
that these vacuum mode processes can be enhanced if collective
excitations of the ensembles are used. As the dynamics takes place,
the two molecular ensembles becomes entangled as the molecules are
effectively coupled through the exchange of energy via the vacuum
mode of the cavity. We will show that this entanglement can be
generated in a comparatively short time. This result provides a
novel route to entangle two mesoscopic systems. Relaxation and
decoherence effects should also be modest as the system is prepared
in a vacuum state; a potentially crucial advantage for practical
applications.

In this paper, we suppose the molecular ensembles are placed at a
distance $d$ apart in an effectively one-dimensional resonator as
shown in Fig. \ref{fig1}. The two states, produced by the splitting
of a rotational ground level by hyperfine interactions are chosen as
the internal molecular states \cite{Rabl}.  They have an energy
difference $\omega$, and are coupled to a single-mode microwave
field with a frequency $\omega_0$. Each of the molecule can be
described by a spin-half particle $\vec{\sigma}_i$, and hence a
collection of such spin-half particles can be described by a
collective angular momentum operator
$\vec{J}=\sum^N_{i=1}\vec{\sigma}_i$, when $N$ is the number of
molecules in one of the ensembles.  The wavelength of microwave
radiation will be much longer than the size of molecular ensemble.
Hence, we can then assume the microwave field coupling to all
molecules for the system of coupled molecules and radiation field in
the form. We can now write down the Hamiltonian $H$ ($\hbar=1$)
\cite{remark}:
\begin{eqnarray}
H&=&\sum^2_{i=1}\omega_0a^{\dag}a+\omega{J}_{z_i}+g_i(a+a^\dag)(J_{+_i}+J_{-_i}),
\end{eqnarray}
Here, $a^\dag$ and $a$ is the creation and annihilator operators of
the cavity mode, $J_{z_i}$ and $J_{x_i}$ are the angular momentum
operators to describe the collective inversion and transition for
the $i$-th ensemble respectively, and $i=1,2$. The molecule-photon
interaction strength is denoted by $g_i$ for the $i$-th ensemble and
they differ with a relative phase $\phi=2\pi\omega{d}/c$ between the
field and two ensembles, where $c$ is the speed of the microwave
photon.  For simplicity, the magnitude for two Rabi coupling
strengths are chosen to be the same, $|g_1|=|g_2|$, and
$\phi{\approx}0$. We consider the case where the molecules and
photon field are in resonance, i.e. $\omega=\omega_0$, which is the
optimal condition to observe the effect of the small vacuum
fluctuations. We note that the Hamiltonian $H$ has the same form as
the Hamiltonian of the Dicke model without the rotating wave
approximation \cite{Lambert}. The analysis we present here applies
to thermal atomic ensembles as well as condensates \cite{Treutlein}.
Condensates would have the advantage of longer coherence times but
also introduce nonlinear dynamical problems \cite{Law}.

\begin{figure}[ht]
\includegraphics[height=5cm]{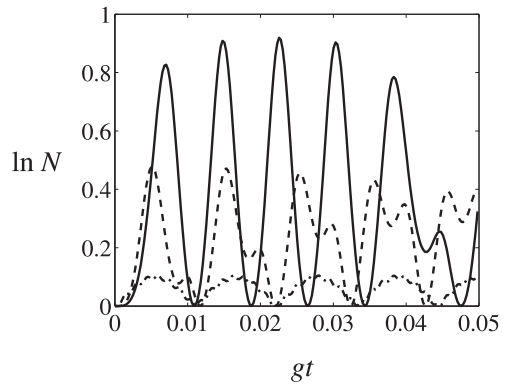}
\caption{\label{fig2} Time evolution (the dimensionless time $gt$)
of the logarithmic negativity ${\ln}N$ with the number of atoms $N$
being $10^4$. The different cases of $\omega=300g$, $500g$ and
$2000g$ are shown in solid, dashed and dash-dotted lines
respectively.}
\end{figure}

To study the quantum dynamics of this system, we need to find the
eigen-states of the whole system. The Hilbert space of this system
is extremely large as the number of molecules are correspondingly
large.  We can, however, consider the time evolution case where
involves only the low-lying excitations of the molecules. We can
make this approximation based on the Holstein-Primakoff
transformation (HPT) \cite{Holstein} which enables us to express
angular momentum operators in terms of harmonic oscillators. In this
manner, we can take the leading approximation and map an angular
momentum operator into a harmonic oscillators by taking the lowest
order version of the HPT \cite{Ng}.  We then get
$J_{x_i}\approx\sqrt{{\omega}N_i}x_{a_i}/\sqrt{2}$,
$J_{y_i}\approx-\sqrt{N_i}p_{a_i}/\sqrt{2\omega}$,
$J_{z_i}\approx(p^2_{a_i}+\omega^2x^2_{a_i})/2\omega-N_i/2$ for
$x_{a_i}$ and $p_{a_i}$ are the position and momentum operators, and
$N_i$ is the number of molecules in the cavity. This approximation
is valid as long as
$\langle{p^2_{a_i}+\omega^2x^2_{a_i}}(t)\rangle/\ll2\omega{N_i}$
\cite{Ng}. It should be a very good approximation for the number of
molecules are sufficiently large \cite{Rabl}.

For convenience, we write the cavity field operators in the
phase-space representation: $x_{c}=(a^\dag+a)/{\sqrt{2\omega_0}}$
and $p_{c}=i\sqrt{{\omega_0}}(a^\dag-a)/{\sqrt{2}}$. We represent
the system in terms of position and momentum operators.  The
Hamiltonian of system can then be rewritten in the form:
\begin{equation}
H'=\frac{1}{2}\sum^2_{i=1}(p^2_{c}+p^2_{a_i}+\omega^2_0x^2_{c}+{\omega^2}x^2_{a_i}
+4g_i\sqrt{{N_i}{\omega\omega_0}}x_{a_i}x_{c}).
\end{equation}
We now want to find the dynamics as the ensembles interact and
become entangled. This problem is clearly related to what of finding
the entanglement between two harmonic oscillators in an open-ended
harmonic chain \cite{Plenio}.

This harmonic system will be in a Gaussian state, which allows us to
quantify the general bipartite entanglement of the system.  The
density matrix of a Gaussian state can be completely determined by
the second-order moments of the position and momentum operators of
the system. We just need to study the reduced density matrix of the
molecular part to find out the entanglement produced between the two
ensembles. This reduced density matrix can be obtained by the
tracing out the cavity mode. The reduced density matrix
$\rho_{a_{1,2}}$ with matrix element
$\langle{X_iX_j+X_jX_i}\rangle-2\langle{X_i}\rangle\langle{X_j}\rangle$,
where $X_i=x_{a_i}$ or $p_{a_i}$.  A quantitative measure of
entanglement can be obtained by using the logarithmic negativity
\cite{Vidal} which gives us an upper bound for distillable
entanglement. The logarithmic negativity in a Gaussian state can be
found as \cite{Vidal}
\begin{eqnarray}
{\ln}N=-\sum_j\log_2[{\rm min}(1,|\gamma_j|)],
\end{eqnarray}
where $\gamma_j$ are the symplectic eigenvalues of the matrix
$\rho_{a_{1,2}}$.

We are now ready to investigate the entanglement dynamics of this
system. We consider the initial state as the state of molecules and
cavity, i.e., the state of the decoupled harmonic oscillators. In
Fig. \ref{fig2}, we plot the time evolution of the entanglement
between the ensembles. The system begins with a separable states and
then the entanglement grows rapidly.  In fact, the quantum state of
two ensembles oscillates between being separable and entangled. This
is very similar to the entanglement propagation in a harmonic chain
in which the two oscillators are periodically entangled and
unentangled \cite{Plenio}.

Moreover, the system achieves the first maximal entanglement within
the time interval $t^*=5^{-3}g^{-1}$. We can estimate this time
$t^*$ with the realistic parameters. If we take $g$ as 1 MHz
\cite{Rabl}, nearly maximal entanglement can be yielded within 5 ns.
This means that a significant entanglement can be obtained rather
quickly. Moreover, no further adjustment of the experimental
parameters or making conditional measurements are required
\cite{Duan1}. The time scale of this entanglement generation ($\sim$
1 ns) is much shorter than the other decoherence sources such as
inelastic collisions ($\sim$ 10 $\mu$s) and photon losses ($\sim$ 1
$\mu$s) \cite{Rabl}. Entanglement can therefore be observed before
decoherence effect set in showing in a natural and efficient way to
generate quantum entanglement for two mesoscopic systems.  In
addition, we can see that a larger ratio of $g/\omega$ can produce a
larger degree of entanglement in Fig. \ref{fig2}, clearly indicating
that counter-rotating terms cannot be neglected in this strong
coupling limit.

\begin{figure}[ht]
\includegraphics[height=5cm]{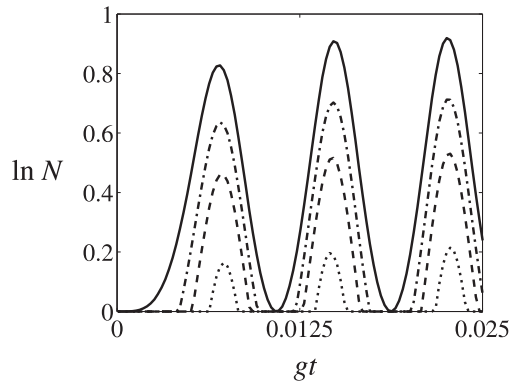}
\caption{\label{fig3} The logarithmic negativity $\ln{N}$ is plotted
against the dimensionless time $gt$ with $\omega=300g$ and $N=10^4$.
The solid, dash-dotted, dashed and dotted lines are represented
$\bar{n}=$ 0, 0.05, 0.1 and 0.2 respectively.}
\end{figure}

We should note that thermal noise is the main potentially problem in
entanglement production.  It is of course impossible to prepare the
perfect vacuum state of a molecular ensemble in an experiment due to
finite temperature effects.  We now assume these ensembles can be
described as a thermal state with mean excitation number
$\bar{n}=[\exp{(\hbar\omega/k_BT)}-1]^{-1}$, for $k_B$ is the
Boltzmann constant and $T$ is the temperature.  We can estimate
$\bar{n}$ to be of order 0.1 to 1 when $\omega\sim${1} GHz and
$T\sim{1}$ to 10 mK \cite{Rabl}.  From this estimation, we can see
that thermal effects cannot be neglected, and it is important to
study their influence on entanglement.  Time evolution of
entanglement under the thermal effects is shown in Fig. \ref{fig3}.
The amount of entanglement produced is shown to be lesser in the
cases of higher $\bar{n}$. Moreover, the longer onset time of
entanglement is required as shown in the higher temperature cases.
But the entanglement can still be observed even if $\bar{n}$ is as
high as 0.2.  This result shows that a substantial amount of quantum
entanglement can be effectively produced using thermal ensembles but
colder molecules due result in a much better performance.

Having discussed the production of entanglement, we now study how to
observe the quantum correlations.  In this Gaussian system, the
density matrix can be constructed if the uncertainties of these two
ensembles can be obtained.  This means that the entanglement of the
two molecular ensembles can be determined just from the quantum
uncertainties. In fact, non-resonant stimulated Raman scattering has
been used to generate and verify the entanglement between two atomic
ensembles \cite{Duan1,ji,Nunn}.   In this scheme, the Stokes pulses
are used to ``write'' the quantum information on the atomic
ensembles and the scattered Stokes photons carry the information of
excitations of each ensembles.  Then, the two Stokes photon fields
coming from each ensemble pass through a 50:50 beam splitter (BS) so
that the two modes interfere and mix together. The conditional
measurement of the resultant Stokes field can be preformed and
entangle the two atomic ensembles \cite{Duan1,ji}. Similarly, the
anti-Stokes pulses can be applied to read the excitations of the
atoms and then the entanglement can be verified by measuring the
correlations of photon fields.

\begin{figure}[ht]
\includegraphics[height=7cm]{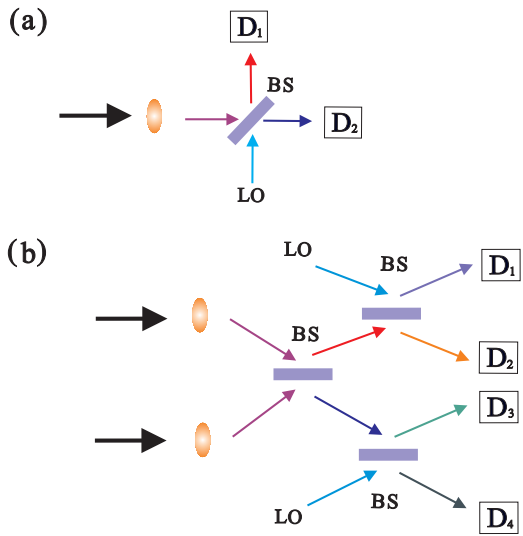}
\caption{\label{fig4} In (a), Illustration for homedyne detection of
measuring quantum correlations of a single molecular ensemble. Two
anti-Stokes pump pulses are applied onto two molecular ensembles.
The output field from molecules is superimposed on the field from a
local oscillator (LO) at a 50:50 BS and then they are detected by
the detectors D${}_1$ and D${}_2$ respectively. In (b), The output
fields from two molecular ensembles are interfered with a 50:50 BS
and then the fields are probed by the balanced homodyne detection.}
\end{figure}

We now adopt this method to determine the entanglement of the two
molecular ensembles as shown in Fig. \ref{fig4}.  We can apply two
anti-Stokes pulses on these two ensembles (being vertical to the
view of Fig. \ref{fig1}) by passing through two small pinholes of
the cavity.  These would have to be larger when the wavelength of
the optical radiation used for the probe and therefore should not
affect the quality of the microwave cavity.  To read out the
excitations of the molecules, we apply an anti-Stokes pulse to each
ensemble which optically pump the excited states to a higher
rotational states with a large detuning between the ground
rotational state. In the Hesienberg picture, the output beam, after
passing through the ensemble \cite{ji}, is given by
$a^{O}_i=\sqrt{\eta_i}{c}_i+\sqrt{1-\eta_i}a^{I}_i$, where $a^{O}_i$
and $a^{I}_i$ are the output field, the vacuum field operator and
$\eta_i$ is the effective transmission efficiency respectively. We
can see that the output field directly carries the information of
collective excitations of the molecules. Hence, the quantum state of
the ensembles can be determined through the measurement of photon
fields.

The measurement of the local and joint quantum correlations of
molecules enables us to determine the logarithmic negativity. We
then require the detection of individual ensemble and the two
ensembles respectively. In Fig \ref{fig4}(a), we give a sketch of a
scheme to measure the local quantum correlations of an individual
ensemble by balanced homodyne detection method via inputting a local
oscillator mode of a coherent state with a large amplitude and phase
$\phi_l$. \cite{Scully}.  The moments $\langle{x^2_i}\rangle$,
$\langle{p^2_i}\rangle$ and $\langle{x_ip_i+p_ix_i}\rangle$ can all
be probed by appropriately adjusting the phase angle $\phi_l$.
Similarly, the joint quantum correlations can also probed by this
method \cite{Scully}. This can be done by interfering two output
fields with a 50:50 BS and then performing balanced homodyne
detection as indicated in Fig. \ref{fig4}(b). The quadrature of the
two modes can be thus determined.

In summary, we have found an efficient method to generate
entanglement between two separate ensembles of molecules and
proposed a method to measure it.  We have assessed the role of the
finite temperature on to the entanglement produced.  It is useful to
the quantum information processing with molecular systems in a
superconducting device. Our study has implication of quantum optics
of mesoscopic system in the strong coupling limit. We envisage that
evaporative cooling of the trapped molecules will realized
\cite{Doyle} so that the temperature can be lowered and the
performance of quantum memory and entanglement generation be further
improved.

H.T.N. thanks the financial support of the Croucher Foundation. K.B.
thanks the Royal Society and Wolfson Foundation for support.


\begin{thebibliography}{99}
\bibitem{Casimir}
H.B.G. Casimir, Proc. Kon. Nederland. Akad. Wetensch. B{\bf 51}, 793
(1948)

\bibitem{Hawking}
S.W. Hawking, Nature {\bf 248} 30 (1974).

\bibitem{Einstein}
A. Einstein, B. Podolsky and N. Rosen, Phys. Rev. {\bf 47}, 777
(1935).

\bibitem{Nielsen}
M. A. Nielsen and I. L. Chuang,  {\it Quantum Computation and
Quantum Information} (Cambridge University Press, Cambridge, 2000).

\bibitem{Reznik}
B. Reznik, Found. Phys. {\bf 33}, 167 (2003).

\bibitem{Retzker}
A. Retzker, J. I. Cirac and B. Reznik, Phys. Rev. Lett. {\bf 94},
050504 (2005).

\bibitem{Cirone}
M. A. Cirone {\it et al.}, Europhys. Lett., {\bf 78}, 30003 (2007).

\bibitem{Andre}
A. Andr\'{e} {\it et al.}, Nat. Phys. 2, 636 (2006).

\bibitem{Rabl}
P. Rabl {\it et al.}, Phys. Rev. Lett. {\bf 97}, 033003 (2006).


\bibitem{Wallraff}
A. Wallraff {\it et al.}, Nature {\bf 431}, 162 (2004); A. Blais
{\it et al.} Phys. Rev. A {\bf 69}, 062320 (2004).


\bibitem{remark1}
The effect of vacuum fluctuations is of course observed in static
phenomena such as the Lamb shift.

\bibitem{Scully}
M. O. Scully and M. S. Zubairy, {\it Quantum Optics} (Cambridge
University Press, Cambridge, 1997).


\bibitem{remark}
Here the static dipolar interactions between the molecules are
neglected because they are very small compared to the strength of
molecule-field interaction.

\bibitem{Lambert}
N. Lammbert, C. Emary and T. Brandes, Phys. Rev. Lett. {\bf 92},
073602 (2004).

\bibitem{Treutlein}
P. Treutlein {\it et al.}, quant-ph/0703199.

\bibitem{Law}
C. K. Law, H. T. Ng and P. T. Leung, Phys. Rev. A {\bf 63}, 055601,
(2001).

\bibitem{Holstein}
T. Holstein and H. Primakoff, Phys. Rev. {\bf 58}, 1098 (1949).

\bibitem{Ng}
H. T. Ng and P. T. Leung, Phys. Rev. A {\bf 71}, 013601 (2005).


\bibitem{Plenio}
M. B. Plenio, J. Hartley and J. Eisert, New J. Phys. {\bf 6}, 36
(2004).

\bibitem{Vidal}
G. Vidal and R. F. Werner, Phys. Rev. A {\bf 65}, 032314 (2002).

\bibitem{Duan1}
L.-M. Duan, M. D. Lukin, J. I. Cirac and P. Zoller, Nature {\bf
414}, 413 (2001).


\bibitem{ji}
W. Ji, C. Wu, S. J. van Enk and M. G. Raymer, Phys. Rev. A {\bf 75},
052305 (2007).

\bibitem{Nunn}
J. Nunn {\it et al.}, Phys. Rev. A {\bf 75}, 011401(R) (2007).

\bibitem{Doyle}
J. Doyle {\it et al.}, Eur. Phys. J. D {\bf 31}, 149–164 (2004).

\end{thebibliography}
\end{document}